\documentstyle[aps,multicol,tabularx,epsf]{revtex} 
\def\d{{\rm d}}
\begin{document}
\draft
\title{Statistical mechanics of semiflexible ribbon polymers}
\author{Ramin Golestanian$^{1,2,4}$ and Tanniemola~B. Liverpool$^{1,3,}$\footnote{Present address: Blackett Laboratory, Imperial College, 
     Prince Consort Road, London SW7 2BZ}}
\address{$^{1}$Max-Planck-Institut f{\"u}r Polymerforschung, D-55021 Mainz, Germany \\ 
$^{2}$Institute for Advanced Studies in Basic Sciences, Zanjan 45195-159, Iran \\
$^{3}$Physico-Chimie Th{\'e}orique, ESA CNRS 7083, E. S. P. C. I., 
75231 Paris Cedex 05, France  \\
$^{4}$Institute for Theoretical Physics, University of California, 
Santa Barbara, CA 93106-4030}
\date{\today}
\maketitle
\begin{abstract}
The statistical mechanics of a ribbon polymer made up of two
semiflexible chains is studied using both analytical techniques and
simulation.  The system is found to have a crossover transition at
some finite temperature, from a type of short range order to a
fundamentally different sort of short range order. In the high
temperature regime, the 2-point correlation functions of the object
are identical to worm-like chains, while in the low temperature regime
they are different due to a twist structure. The crossover happens
when the persistence length of individual strands becomes comparable
to the thickness of the ribbon.  In the low temperature regime, the
ribbon is observed to have a novel ``kink-rod'' structure with a
mutual exclusion of twist and bend in contrast to smooth worm-like
chain behaviour. This is due to its anisotropic rigidity and
corresponds to an {\it infinitely} strong twist-bend coupling. The
double-stranded polymer is also studied in a confined geometry. It is
shown that when the polymer is restricted in a particular direction to
a size less than the bare persistence length of the individual
strands, it develops zigzag conformations which are indicated by an
oscillatory tangent-tangent correlation function in the direction of
confinement. Increasing the separation of the confining plates leads
to a crossover to the free behaviour, which takes place at separations
close to the bare persistence length.  These results are expected to
be relevant for experiments which involve complexation of two or more
stiff or semiflexible polymers.
\end{abstract}
\pacs{87.15By, 36.20Ey, 61.25Hq}

\section{Introduction and Summary}      \label{sIntro}

There has been a lot of recent interest in physical properties of
biopolymers, ranging from elasticity of biopolymer networks and 
its use in the prediction of mechanical properties of cells to  
direct visualisation of single chain properties. 
Examples of important biological macromolecules whose
physical properties have been recently studied are: actin, a
double-stranded semiflexible protein polymer which forms an integral
part of the cytoskeleton (mechanical structure) of eukaryotic (e.g.
fungi, plants, animals) cells; microtubules, multi-stranded rigid and
dynamic protein polymers which form one of the main components of the
cytoskeleton of eukaryotic cells and play an important part in their
organisation; DNA, which carries the genetic code of all living
organisms~\cite{bio}.  

Since many of the processes involved in cell function(e.g. DNA
replication in cell division) require major structural changes of
these biopolymers~\cite{strick}, there is a need for more microscopic
but still analytically tractable models of such polymers which go
beyond the simple picture of such molecules as homogenous elastic
rods~\cite{homo}.  Motivated by this, we study such a microscopic
model specifically looking for qualitative differences between the
behaviour of such molecules and simple worm-like
chains~\cite{KratPor}.  In addition, most analyses of the worm-like
chain models of polymers have focused on ground state properties
(long chains) or bulk quantities~\cite{Odijk}.  It is interesting to
look at the effect of fluctuations, spatial correlations {\it and}
finite size on these systems.

A double-stranded semi-flexible polymer chain is the basic structure of many
biopolymers. Examples are of double-stranded biopolymers are DNA and proteins
such as actin. The model most used in the study of biopolymers is that of
the worm-like chain~\cite{KratPor} in which the polymer flexibility
(structure) is determined by a single length, the persistence length
$\ell_{p}$ which measures the tangent--tangent correlations.
For example, DNA has a persistence length $\ell_{p} \approx
50\mbox{nm}$ whilst for actin $\ell_{p} \approx 17\mu\mbox{m}$. These
biopolymers are known to have a more complex `twisted' structure.
The multi-stranded nature of these polymers is
also not taken into account in a simple worm-like chain model.
It is not clear if such a fine structure will have an effect on the global
properties of these objects. A possible effect of
such fine structure is what we attempt to study in this article. Our model is,
in a sense, microscopic because the interaction between the bend and twist
degrees of freedom is a {\it result}. This is fundamentally different from
previous approaches~\cite{homo} which try to include the twist
degrees of freedom by adding extra terms to the free energy.

In a previous letter~\cite{LGK} we studied a version of the rail-way
track model of Everaers-Bundschuh-Kremer (EBK) \cite{EBK} for a
double-stranded semi-flexible polymer, embedded in a $d$-dimensional
space for arbitrary $d$. The main purpose of this article is to
present a detailed description of our theoretical and numerical
calculations and in addition we present some new results on the effect
of confinement on the statistical mechanics of the ribbon polymer \cite{china}.
Excluded volume and electrostatic interactions have been ignored
throughout.

We find that the system, has qualitatively different properties in the
low temperature and high temperature regimes, in contrast to what one
might naively expect from an inherently one-dimensional system with
local interactions and constraints. The tangent--tangent correlation
function decays exponentially in the whole range of temperatures with
a ``tangent-persistence length'' ${\ell}_{\rm TP}$ that has a
very slow temperature dependence, and whose scale is determined by the
(bare) persistence length of a single strand $\ell_p = \kappa/k_{B}T$ ($\kappa$
is the bending stiffness of a single strand). Note that it is
independent of $a$, the separation of the two strands, which is the
other relevant length scale in the problem.  However, the correlation
function of the bond-director field, defined as a vector that
determines the separation and coupling of the two strands of the
combined polymer system, has different behaviour below and above the
temperature $T_{\times} \simeq 4.27 \kappa/d k_{B} a$.  While it decays purely
exponentially for $T > T_{\times}$, there are additional oscillatory
modulations for $T < T_{\times}$. The related ``bond-persistence length''
${\ell}_{\rm BP}$ does not change appreciably at high
temperatures, where its scale is again set by $\ell_p$ alone.  In the
low temperature phase, however, ${\ell}_{\rm BP}$ does show a
temperature dependence. In particular, ${\ell}_{\rm BP} \sim
\ell_{p}^{1/3} a^{2/3} \propto T^{{-1/3}}$ for $T \sim 0$, while
${\ell}_{\rm BP} \sim \ell_{p}$ for $T \sim T_{\times}$. Similarly, the
``pitch'' $H$, defined as the period of oscillations in the low
temperature regime, changes drastically with temperature , ranging
from $H \sim {\ell}_{\rm BP} \sim \ell_{p}^{1/3} a^{2/3}$ near $T=0$,
to $H \sim 0$ near $T=T_{\times}$.  At $T=0$ we restore a flat ribbon which
has true long-range order in both the tangent and bond-director
fields. The ribbon is essentially a rigid rod. As we approach $T=0$,
the persistence lengths and the pitch diverge with the scaling $H \sim
{\ell}_{\rm BP} \sim {\ell}_{\rm TP}^{1/3}$.

The spontaneous appearance of a short range twist structure may be
understood in the language of the {\em homogenous} rod
models~\cite{homo} as a {\em local} twist-bend coupling, which is
observed up to a screening length $\ell_{\rm BP}$ that lies between
$a$ and $\ell_p$.  We find that the anisotropy in the rigidity of a
ribbon results in a ``kink-rod'' structure, in which the ribbon is
{\em at every point along its contour} either twisted and unbent (rod)
or bent and untwisted (kink).  This {\em inhomogeneous} behaviour is in
sharp contrast with uniform worm-like chain behaviour, and can be
interpreted as an {\it infinitely} strong twist-bend coupling. This
structure is, however, screened on long length scales due to the fact
that the ribbon is a one dimensional system with short range
interactions. The short range twist order and the kink-rod structure
will naturally disappear when $a \sim \ell_p$ corresponding to the
temperature $T_{\times}$. We also observe a twist-stretch coupling.

We also study the effects of confinement on the double-stranded
polymer.  We enforce the confinement to a box of size $R$ as an
additional constraint. For $R \gg \ell_p$, we recover the behaviour of
free double-stranded semiflexible polymers \cite{LGK}. In particular,
the tangent-tangent correlation has a purely exponential decay with a
characteristic length scale of order $\ell_p$, while the bond-director
field develops a crossover to a phase with oscillatory correlations.
As $R$ is decreased, there is a crossover to a phase with oscillatory
tangent-tangent correlations about $R \sim \ell_p$.  For $R \ll \ell_p$,
we find that both the persistence length in the perpendicular
direction $\ell_{\perp}$, and the characteristic oscillation length
$\lambda$, scale as $(\ell_p R^2)^{1/3}$ which is considerably less than
$\ell_p$.  The same crossover in the bond-director field also persists
in this limit, and in particular, there is a regime in which both the
tangent and the bond-director correlations are oscillatory.

We introduce and define our model in Sec. \ref{sRail}, and then
describe a mean field approach in Sec. \ref{sFree}, which can be used
to obtain closed form expressions for various correlation
functions. In Sec. \ref{sPlaq} we discuss a physical argument for the
results we obtain using a plaquette model. In Sec. \ref{sSimul} we
describe some extensive Molecular Dynamics/Monte Carlo simulations,
which we use to calculate the correlation functions, and compare them
with the mean field results of Sec. \ref{sFree}. We discuss the thermal
kink-rod structure of stiff ribbons in Sec. \ref{sKink}, and the effects of
confinement on the conformations of ribbon polymers in Sec.
\ref{sConfine}. Finally, Sec. \ref{sConcl} summarises our results and
the limitations of our approach.
  
\section{Railway Track Model}           \label{sRail}

\begin{figure}
\centerline{\epsfxsize 6cm \epsffile{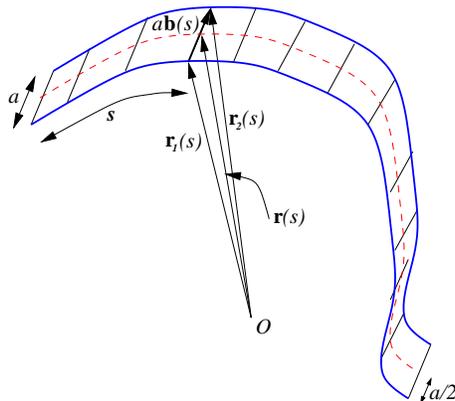}}
\caption{The schematic of the double-stranded semiflexible polymer of 
two chains separated by a distance $a$. Note the bond director field ${\bf
    b}(s)$.}
\label{fig1}
\end{figure}

To study the effect of a double-structure on semiflexible polymers, we
consider a version of the railway track model of
Everaers-Bundschuh-Kremer (EBK) \cite{EBK}.  In our approach, we are
able to consider polymers embedded in a $d$-dimensional space for
arbitrary $d$.  The system is composed of two semiflexible chains,
each with rigidity $\kappa$, whose embeddings in $d$-dimensional space are
defined by ${\bf r}_1 (s)$ and ${\bf r}_2 (s)$.  The Hamiltonian of
the system can be written as the sum of the Hamiltonians of two
wormlike chains
\begin{equation}
{\cal H}=\frac{\kappa}{2} \; \int \d s \; \left[
\left(\frac{\d^2 {\bf r}_1 (s)}{\d s^2} \right)^2
+\left(\frac{\d^2 {\bf r}_2 (s)}{\d s^2} \right)^2 \right].\label{H12}
\end{equation}
We assume that the individual strands 
(that make up the double-stranded polymer) are
inextensible: $(\d {\bf r}_1/\d s)^2=(\d {\bf  r}_2/\d s)^2=1$. 
The ribbon structure is then enforced by having  ${\bf r}_2(s')$ separated from 
${\bf  r}_1(s)$ by a distance $a$, i.e. ${\bf r}_2(s') = {\bf r}_1(s)+ a{\bf 
b}(s)$, 
where $|s-s'|$ can be non-zero but is small. We
have defined a {\it bond-director field} ${\bf  b}(s)$, which is a unit vector
perpendicular to {\it both} strands (see Fig.~\ref{fig1}).  The chains are
assumed to have permanent bonds (such as hydrogen bonds) that are strong enough
to keep the distance between the two strands constant. In Ref.\cite{EBK},
it is argued that the relevant constraint on the system would then 
require that the arclength mismatch between the two strands in a bent 
configuration should be very small. We can calculate the
arclength  mismatch for the bent configuration as
$ \Delta s=|{\bf r}_2 (s)-{\bf r}_1 (s)+a \;{\bf b}(s)| $,
where $a$ is the separation of the strands. We impose the constraint
as a hard one, namely, we set $\Delta s =0$ as opposed to Ref.\cite{EBK}.
Physically this means we {\it do not} allow bends in the plane of the ribbon.
These bends are not important in $d>2$ because as we shall see the lower
length scale will be set by the `pitch' which will make the in-plane
fluctuations of the ribbon irrelevant~\cite{EBK}.

We can argue that the simplifying assumption does not change the
behaviour of the system.  If we impose a soft constraint as in
Ref.\cite{EBK} using an energy term like $(k/2) \int (\Delta s)^2$, we can see
that the length $l=(\kappa/k a^2)^{1/2}$ determines two different regimes;
the interesting one being $L \gg l$ ($L$ is the length of the chains).
Hence, our hard constraint in fact only restricts us to the case of
interest.

We implement the constraint $\Delta s =0$ by introducing the ``mid-curve''
${\bf r}(s)$:
\begin{eqnarray}
{\bf r}_1 (s) &=& {\bf r}(s)+\frac{a}{2} \; {\bf b},\label{r1r2} \\
{\bf r}_2 (s) &=& {\bf r}(s)-\frac{a}{2} \; {\bf b}.\nonumber
\end{eqnarray}
In terms of the tangent to the mid-curve ${\bf t}=\d {\bf r}/\d s$,
which we call the {\it tangent-director field}, and the bond-director 
${\bf b}$, the Hamiltonian of the system can now be written as
\begin{equation}
{\cal H}=\frac{\kappa}{2} \; \int \d s \; \left[
2 \left(\frac{\d {\bf t} (s)}{\d s} \right)^2
+\frac{a^2}{2}\left(\frac{\d^2 {\bf b}(s)}{\d s^2} \right)^2
\right],\label{Htn}
\end{equation}
subject to the exact (local) constraints
\begin{eqnarray}
({\bf t}\pm\frac{a}{2} \;\frac{\d {\bf b}}{\d s})^2=1, \qquad
{\bf b}^2=1,\label{constr} \\
({\bf t}\pm\frac{a}{2} \;\frac{\d {\bf b}}{\d s}) \cdot {\bf b}=0, \qquad
\nonumber
\end{eqnarray}
For a weakly bent ribbon, the Hamiltonian in Eq.(\ref{Htn}) can be
conveniently thought of as having two major contributions:
a bending energy ${\cal H}_{b}$ (the first term), and  
a twisting energy ${\cal H}_{t}$ (the second term).

\section{Mean Field Theory: Free Chains}        \label{sFree}

It is well known that the statistical mechanics of semiflexible chains
are difficult due to the constraint of inextensibility. Various
approximation methods have been devised to tackle the problem. A
successful scheme, that somehow manages to capture the crucial
features of the problem, is to impose global (average) constraints
rather than local (exact) ones
\cite{Bawendi,Lagowski,Gupta,Liverpool,Ha}. This approximation is
known to be good for calculating the average end-to-end length. It can
be used for the probability distribution of the end-to-end length {\it
  only} if the persistence length is much less than the chain length
so that the chain conformation can be considered isotropic.  We can
get good insight to the approximation scheme, by considering the fact
that it corresponds to a saddle-point evaluation of the integrals over
the Lagrange multipliers, which are introduced to implement the
constraints \cite{Liverpool,Ha}. In this sense, it is known to be a
``mean-field'' approximation in spirit. One can then go further by
considering the effect of fluctuations on this mean-field result. The
power of this approach is that one can easily calculate quantities
which turn out to be very difficult if the constraints are required to
hold exactly~\cite{Odijk}.

To study the effect of fluctuations, we have performed a
systematic $1/d$-expansion (sketched in Appendix \ref{a1/d})\cite{David}. 
We see that no divergent behaviour appears when we calculate the diagrams of the
{\it 2-point} correlation functions. This means that the mean-field
behaviour of these functions, at least, won't change due to
fluctuations, although it does not preclude differences in higher order
correlation functions.

With the above discussion as justification, we apply the same approximation
scheme to our problem defined in Sec.~\ref{sRail}: The local
constraints in Eq.(\ref{constr}) are relaxed to global ones. This can be
done by adding the corresponding ``mass terms'' to the Hamiltonian
\begin{eqnarray}
\frac{{\cal H}_{m}}{k_{B}T} = \int \d s \left[
\frac{b}{\ell_p}\left({\bf t}-\frac{a}{2} \;\frac{\d {\bf b}}{\d s}\right)^2
+\frac{b}{\ell_p}\left({\bf t}+\frac{a}{2} \;\frac{\d {\bf b}}{\d s}\right)^2
+\frac{c a^2}{4 \ell_{p}^3} {\bf b}^2 \hskip 2cm \right. \label{Hm} \\
\hskip 3cm
+\left.\frac{e}{\ell_p}
\left({\bf t}-\frac{a}{2} \;\frac{\d {\bf b}}{\d s}\right) \cdot {\bf b}
+\frac{e}{\ell_p}
\left({\bf t}+\frac{a}{2} \;\frac{\d {\bf b}}{\d s}\right) \cdot {\bf b}
\right],\nonumber
\end{eqnarray}
where $b$, $c$, and $e$ are dimensionless constants. The partition function
is then given by
\begin{equation}
Z[{\bf J},{\bf K}]= \int {\cal D}{\bf t}(s)\;{\cal D}{\bf b}(s) 
\exp\left\{{-\frac{{\cal H}+{\cal H}_{m}}{k_{B}T}
+\int \d s \left[{\bf J}(s)\cdot{\bf t}(s)+{\bf K} (s)\cdot{\bf b}(s)
\right] }\right\}
\end{equation}

\begin{figure}  
\centerline{\epsfxsize 10.0cm \epsffile{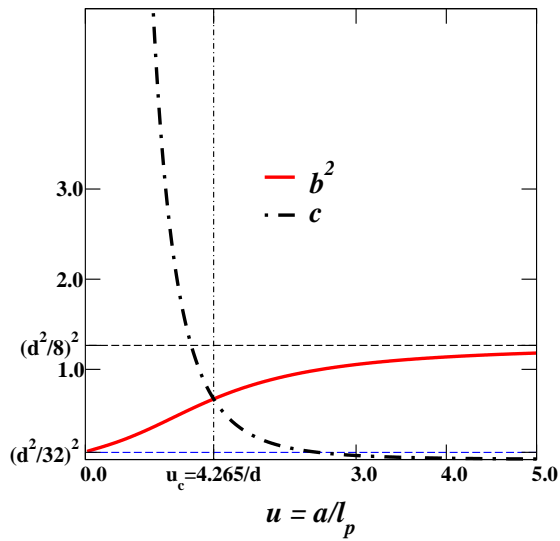}}
\caption{The solution of the self consistent equations for the
constants $b$ and $c$ as a function of $u = a/\ell_{p}$ in $d=3$. The
value $u_{c} \simeq 4.27/d$ corresponds to the transition point.}
\label{solution}\end{figure}

We next determine the constants self consistently by demanding the
constraints of Eq.(\ref{constr}) to hold on average, where the thermal
average is calculated by using the total Hamiltonian ${\cal H}+{\cal
  H}_m$.  Note that in choosing the above form, we have implemented
the ``label symmetry'' of the chains, namely, that there is no
difference between two chains.  It is convenient to take the limit of
an infinitely long chain and perform the functional integrals in
momentum space. The Fourier transforms are defined $\tilde{\bf A}(q) =
\int \d s \exp \{ i q s \} {\bf A}(s)$. We have
\begin{equation}
Z[\tilde{\bf J},\tilde{\bf K}]= \int {\cal D}\tilde{\bf t}(q)\;
{\cal D}\tilde{\bf b}(q) \exp \left\{ - {1 \over 2}\int {\d q \over 2 \pi} 
(\tilde{\bf t}(-q),\tilde{\bf b}(-q)) \cdot \bar{\bf M}(q) \cdot 
\left(\begin{array}{c} \tilde{\bf t} (q)\\ \tilde{\bf b} (q) \end{array}
\right) +\int {\d q \over 2 \pi} \left[\tilde{\bf J} \cdot\tilde{\bf t}+ 
\tilde{\bf K} \cdot \tilde{\bf b}\right] \right\}.
\end{equation}
The Gaussian integration can then be easily performed. It yields 
\begin{equation}
Z[\tilde{\bf J},\tilde{\bf K}]= \exp\left\{  {1 \over 2} 
\int {\d q \over 2 \pi} (\tilde{\bf J}(-q),\tilde{\bf K}(-q)) \cdot \bar{\bf M}^{-1}(q) \cdot 
\left(\begin{array}{c} \tilde{\bf J} (q)\\ \tilde{\bf K} (q) \end{array}
\right) \right\},
\end{equation}
where 
\begin{equation}
\bar{\bf M}(q) = \left[ \begin{array}{cc} 2 \ell_p q^2 + {4 b /\ell_p} & 
{2 e / \ell_p} \\  \\
{2 e / \ell_p} & {\ell_p a^2  q^4 / 2} + {b a^2 q^2 / \ell_p}  
+ { c a^2 / \ell_p^3}\end{array}\right].
\end{equation}
The averages are easily obtained from $Z$, for example
$$\langle \tilde{t}_i(q) \cdot\tilde{b}_j(q') \rangle  = 
{\delta^2 \ln Z \over \delta \tilde{J}_i (q) \delta \tilde{K}_j (q') }$$
The next step is to demand self consistently that 
\begin{eqnarray}\left \langle {\bf b}(s)^2\right\rangle &=& 1 \,  ,
 \nonumber \\ 
\left\langle \left({\bf t}(s)\pm\frac{a}{2} \;\frac{\d {\bf b}(s)}{\d s}\right)^2\right\rangle 
&=& 1 \, , \label{selfcons} \\
\left\langle \left({\bf t}(s) \pm\frac{a}{2} \;\frac{\d {\bf b}(s)}{\d s}\right) \cdot 
{\bf b}(s) \right\rangle &=& 0 \, , \nonumber 
\end{eqnarray}
The self-consistency leads to the following set of
equations for the constants $b$, $c$ and $e$:
\begin{equation}        \label{bceqn}
\left\{ \begin{array}{l}
\frac{1}{4 \sqrt{2 b}}+\frac{a^2 \sqrt{c}}{4 d \ell_{p}^2}=\frac{1}{d} 
\\ \\
c (b+\sqrt{c})=\frac{d^2 \ell_{p}^4}{2 a^4}  \nonumber \\ \\
e=0 \nonumber
\end{array} \right. .
\end{equation}     
The above equations, which are nonlinear and difficult
to solve exactly, determine the behaviour of $b$ and $c$ as a
function of $u=a/\ell_p$. We have solved them numerically in $d=3$
and the solutions are given in Fig.~\ref{solution}.
One can solve Eq.(\ref{bceqn}) analytically
in two limiting cases. For $u \ll 1$ we find $b=d^2/32$ , and
$c=(d/\sqrt{2})^{4/3} u^{-8/3}$, whereas for $u \gg 1$ we find
$b=d^2 /8$, and $c=4/u^4$. In Fig.~\ref{solution}, the behaviour of $b$ and $c$
is plotted as a function of $u$. Note that $u$ is proportional
to $T$ and can be viewed as a measure of temperature.

We can then calculate the correlation functions.
For the tangent--tangent correlation one obtains
\begin{equation}
\left<{\bf t}(s) \cdot {\bf t}(0)\right>=\frac{d}{4 \sqrt{2 b}}
\exp\left(-\sqrt{2 b} \;\frac{s}{\ell_p}\right),\label{tt}
\end{equation}
whereas for the bond-director field one obtains
\begin{equation}
\left<{\bf b}(s) \cdot {\bf b}(0)\right>=
\frac{d \ell_{p}^2}{2 a^2 \sqrt{b^2-c}} \left[
\frac{\exp\left(-(b-\sqrt{b^2-c})^{1/2}\;\frac{s}{\ell_p}\right)}
{(b-\sqrt{b^2-c})^{1/2}}
-\frac{\exp\left(-(b+\sqrt{b^2-c})^{1/2}\;\frac{s}{\ell_p}\right)}
{(b+\sqrt{b^2-c})^{1/2}} \right].\label{nn}
\end{equation}
The tangent--tangent correlation (Eq.(\ref{tt})) is exactly what we
obtain for a single worm-like chain, and implies uniform behaviour for
all temperatures. Eq.(\ref{nn}) on the other hand, indicates a change
of behaviour at $b^2=c$ for the bond-director correlation. The
correlation is {\it over-damped} for $b^2 > c$ (high temperatures),
while it is {\it under-damped} (oscillatory) for $b^2 < c$ (low
temperatures). The interesting point $b^2=c$ happens for
$u_c=16/(1+\sqrt{2})^{3/2} d \simeq 4.27/d$, that leads to the value
for $T_{\times}$ quoted above (see Fig.~\ref{solution}).  We also find
a divergence in the specific heat,
$C_{V}=\frac{\partial^{2}F}{\partial T^{2}}$ where $F=-k_{B}T\log Z$
at $T_{\times}$. It should be noted that it is not a thermodynamic
phase transition in the sense of long-range ordering and broken
symmetry. It is a crossover that appears due to competing effects, and
the transition is from a state with some short-range order to a state
with a different short-range order. Similar phenomena have been
observed in Ising-like spin systems with competing
interactions~\cite{Hornreich} and the crossover (transition) point
corresponds to a type of {\it`Lifshitz point'} for a 1-$d$ system.


It is interesting to study the bond-director correlation in the
limiting case $b^2 \ll c$, that corresponds to relatively low
temperatures. Using the asymptotic forms for $b$ and $c$, one obtains
\begin{equation}
\left<{\bf b}(s) \cdot {\bf b}(0)\right>=
\sqrt{2}\;\exp\left[-\left(\frac{d}{4 \ell_p a^2}\right)^{1/3} s\right]
\;\times\;\sin\left[\left(\frac{d}{4 \ell_p a^2}\right)^{1/3} s+\frac{\pi}{4}
\right],\label{lowTnn}
\end{equation}
for very low temperatures. From the above expressions for the correlation 
functions, one can
read off the persistence lengths ${\ell}_{\rm TP} \sim \ell_p$ and
${\ell}_{\rm BP} \sim (\ell_p a^2)^{1/3}$, and the pitch $H \sim (\ell_p a^2)^{1/3}$.

>From the tangent-tangent correlation function, we can calculate the
end-to-end distance. It yields
\begin{eqnarray}        \label{endtoend}
\left< \left( {\bf r}(s)-{\bf r}(0) \right)^2 \right>&=&
\int_0^s \int_0^s \d s_1 \d s_2 \left< {\bf t}(s_1) \cdot {\bf t}(s_2) \right> \\
&=& {d \ell_p \over 4 b} \left[s- {\ell_p \over \sqrt{2 b}}
\left(1-{\rm e}^{-\sqrt{2 b} s/\ell_p}\right)\right], \nonumber
\end{eqnarray}
which is similar to worm-like chains. It interpolates between the
limiting behaviours of random walks ($\sim (d \ell_p /4 b) s$) for $s
\gg \ell_p$ to rods ($\sim (d/4 \sqrt{2 b}) s^2$) for $s \ll \ell_p$.
However, it is interesting to note that there is a shrinking in the
length of the rod by a factor of $(d/4 \sqrt{2 b})^{1/2}$, which
varies smoothly from $1$ at $a \ll \ell_p$ to $1/\sqrt{2}$ at $a \gg
\ell_p$. This implies that a polymer made up of two inextensible
strands is always ``slightly extensible'' at any finite temperature,
due to the presence of twist fluctuations. This is exactly the {\em
twist-stretch} coupling studied by various authors using homogenous 
elastic rod models ~\cite{homo}. Note that in our model this coupling
is a {\em result}, as opposed to the elastic rod models in which it must 
be added by hand. A microscopic model proposed
by O'Hern et al~\cite{stack}, which describes DNA as a stack 
of plates, also predicts a twist-stretch coupling.

A simple scaling argument can account for $\ell_{\rm TP}$ and
$\ell_{\rm BP}$ in the low temperature regime.  Consider applying a
uniform bend of radius of curvature $\lambda$ to a section of ribbon
of length $\lambda$ without twisting it. The corresponding bending
energy (calculated using ${\cal H}_{b}$) is given by $ {E}_{b} \sim
k_B T \ell_p / \lambda$. We can then estimate $\ell_{\rm TP}$ by
finding the length $\lambda$ for which the bending energy $E_b$
becomes comparable to $k_B T$. Similarly, applying a uniform twist per
length $2 \pi /\lambda$ to a section of the ribbon of length $\lambda$
without bending will cost a twist energy ${E}_{t} = k_B T
\ell_p a^2 / \lambda^3 $ (calculated using ${\cal H}_{t}$). 
The wave-length $\lambda$ at which the twist energy $E_t$ becomes
comparable to $k_B T$ gives $\ell_{\rm BP}$.

\section{Plaquette Model and Competition}       \label{sPlaq}

\begin{figure}
\centerline{\epsfxsize 8.0cm \epsffile{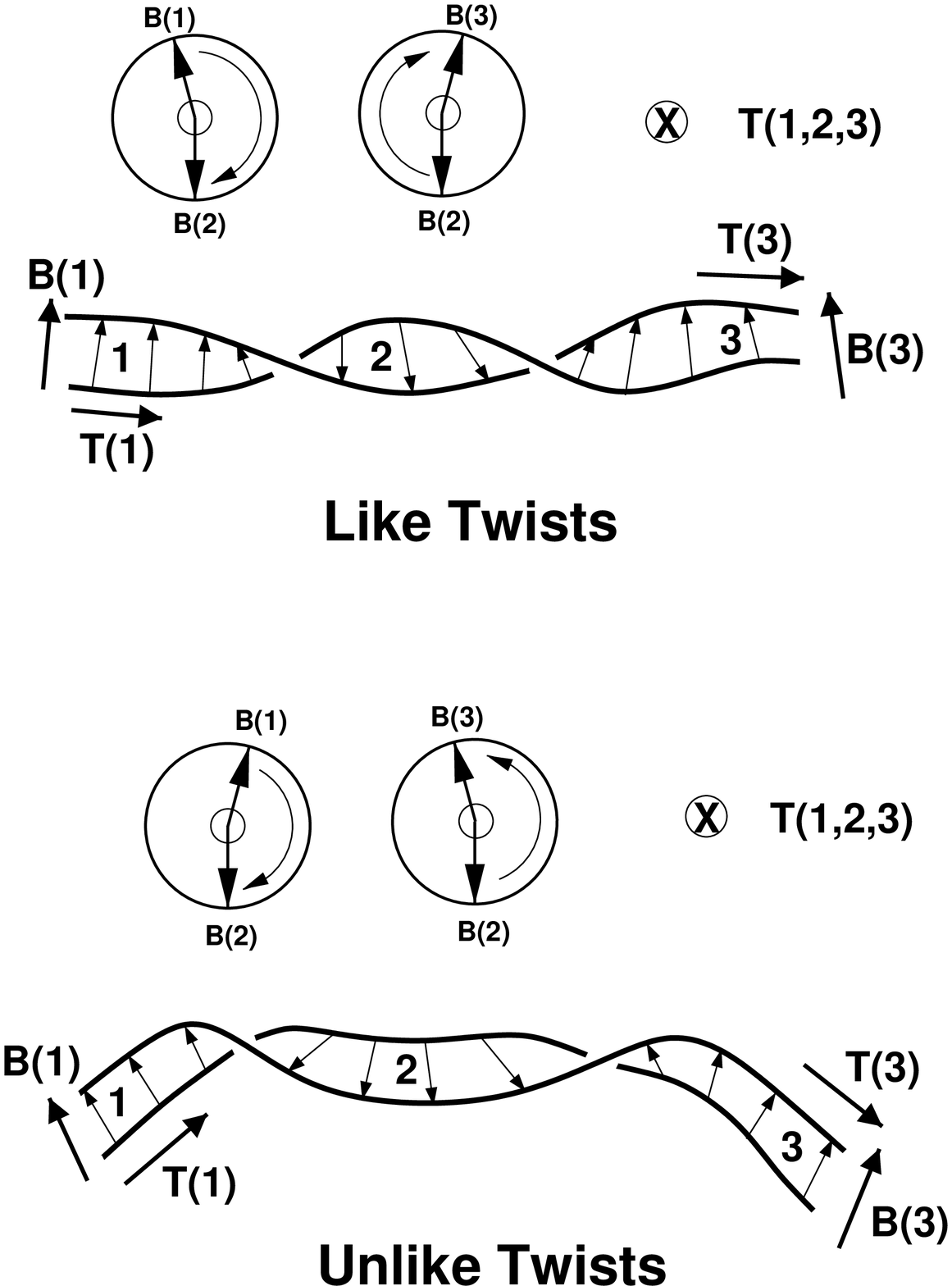}}
\caption{Coarse-grained model - Like and Unlike twists meeting. The bond vector ${\bf B}$ is rotating about the tangent vector in the same direction for like twists and opposite direction for unlike twists.}
\label{fig:liketwists}
\end{figure}

The nature of competition in our double-stranded polymer system can be
understood using a plaquette model.
We can coarse-grain the ribbon to a length-scale ($\ell$) where we can
consider it to be made up of plaquettes which are joined up to form a
ribbon (see Fig. \ref{fig:liketwists}). We can then define effective
coarse-grained bond $\hat{\bf B}_i$ and tangent $\hat{\bf T}_i$
director fields for each plaquette. 

The energy expression corresponding to bends comes from the product of
the tangent-directors of the neighbouring plaquettes:
\begin{equation}
\beta {\cal H}_{b} = -{\ell_p \over
  \ell} \sum_i \hat{\bf T}_i \cdot \hat{\bf T}_{i+1}.
\end{equation}
In a spin analogy, this corresponds to a classical Heisenberg
ferromagnet in 1-$d$, and has no competition .

If we choose $ \ell \ll \ell_p$ in the coarse-graining process, we can
safely assume that the ribbon is rod-like (we freeze out the bending
modes) and that it only has twist fluctuations. We may then write ${\bf
b}(s) = \hat{\bf e}_1 \cos \theta(s) + \hat{\bf e}_2 \sin \theta(s)$
where the (fixed) unit vectors $\hat{\bf e}_i,\{i=1,2\}$ span the
plane perpendicular to the rod. Rewriting Eq.(\ref{Htn}) in terms of
$\theta(s)$ and implementing the constraints, we obtain
\begin{equation}
\beta {\cal H} = {\ell_p a^2 \over 4} \int ds \left [ |\partial_s^2 \theta|^2 
+|\partial_s \theta|^4 + { a^2 |\partial_s^2 \theta|^2 |\partial_s \theta|^2
\over 4 - a^2 |\partial_s \theta|^2}\right], 
\end{equation} 
subject to the constraint: $|\partial_s \theta(s)| < 2 / a$. Note that
the lowest order contribution to the twist potential starts from a
quartic term.

We can now expand the nonlinear term in the above Hamiltonian, and
perform the coarse-graining, in the framework of a perturbation
theory, by integrating out the modes between $1/\ell$ and $1/a$ in the
momentum shell. We can then determine the form of the coarse-grained
Hamiltonian and calculate the renormalized coupling constants. We keep
only terms up to second (Gaussian) order, which is a good
approximation for $\ell < (\ell_p a^2)^{1/3}$, and obtain
\begin{equation}
\beta {\cal H}_{t}= - J_1 \sum_i  \hat{\bf B}_i
\cdot \hat{\bf B}_{i+1} - J_2 \sum_i \hat{\bf B}_i \cdot \hat{\bf B}_{i+2}
\end{equation}
where $$J_1 = c_0 + c_1 {\ell_p a^2 \over \ell^3}, \; \mbox{and} 
\; J_2 = - c_2 {\ell_p a^2 \over \ell^3}\, ,$$ where $c_i, 
\{i=0,\cdots,2\}$ are constants of order unity. In contrast 
to the bending energy, the effective twist energy is frustrated
due to the opposite sign of $J_1$ and $J_2$ ~\cite{Hornreich}. 
In the spin analogy, this 
would correspond to a model with next nearest neighbour competing 
interactions similar to the so-called ANNNI model, which develops 
oscillations for certain values of the ratio $-J_2/J_1$ (of order 1)~\cite{Hornreich}. 
This corresponds to $\ell \sim (\ell_p a^2)^{1/3}$, and we can thus
account for the pitch $H \sim (\ell_p a^2)^{1/3}$.

The competition is present only at nonzero temperatures, and is merely
due to topological constraints of the ribbon. Another, more physical
way of understanding this competition is to consider the interaction
between two neighbouring twisted regions. It is easy to see that
twists of opposite sign meeting at an edge tend to unwind (annihilate)
each other~\cite{like}, while twists of the same sign are trapped when
they meet; they do not annihilate each other and add up (see
Fig. \ref{fig:liketwists}).

\section{Simulation}    \label{sSimul}

An intriguing feature of the behaviour of this model is that, although
the ground state ($T=0$) configuration of the system is a flat ribbon,
and supports no twists, upon raising the temperature, a
twisted structure with short range {\em twist} order develops. We have
confirmed this by performing extensive Molecular Dynamics(MD)/Monte
Carlo (MC) simulations of double-stranded semiflexible polymers. A
bead-spring model with bending and stretching energies was used. We
combined a velocity Verlet MD coupled to a heat bath with an
off-lattice pivot MC algorithm. The MD was useful for equilibrating
the shorter length-scales and MC for the long length-scales.

\begin{figure}
\centerline{\epsfxsize 7.0cm \epsffile{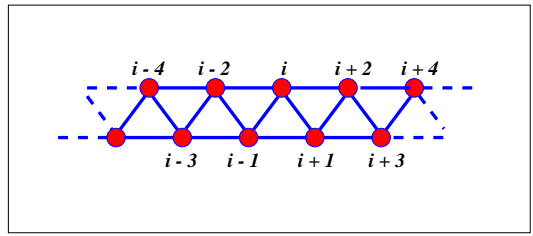}}
\caption{The schematic of the double-strand model used in the simulation.}
\label{fig:simul}
\end{figure}

We used a {\em triangular} lattice to discretise the ribbon (see Fig.
\ref{fig:simul}). The position of the $i$th bead is ${\bf r}_{i}$ and
we assume all the beads have mass $m$. The two chains making up the
double strands join the odd ($\{1,3,5,\ldots,799\}$) and even ($\{2,4,6,\ldots ,
800\}$)beads together.  The Potential Energy is given by
\begin{equation}\frac{U [\{ {\bf r}_i \}]}{k_{B}T}= \sum_{i=1}^{N-2}
    k_{s}[({\bf r}_{i+1}-{\bf r}_{i})^{2}- \ell_0^{2}] + k_{s}[({\bf
      r}_{i+2}-{\bf r}_{i})^{2}- \ell_0^{2}] - k_{b} \cos \theta_{i},
  \end{equation} where
$$\cos \theta_{i} = ({\bf r}_{i+2}-{\bf r}_{i}) \cdot ({\bf r}_{i}-{\bf
  r}_{i-2}).$$
We have a bending constant $k_{b}$ {\em only} for the
springs joining beads on the same chain and a stretching constant $k_{s}$ for
every spring. We also have a short-range repulsion between nearest neighbour
beads.  The MD simulation is performed by integrating a Langevin equation for
every bead
\begin{equation}
m \frac{\d^{2}{\bf r}_{i}}{\d t^{2}} + \Gamma {\d {\bf r}_i \over \d t} 
= - \nabla_{r_{i}}U + f_{i},
\end{equation}
where $f_{i}$ is a random number chosen from a range set by $T$
representing the heat bath. The simulations were performed at
$k_BT=1$. We are in the dissipative regime so we can ignore the
inertial term. The friction term is set to ($\Gamma=0.7$) and the
noise is chosen so as to satisfy the fluctuation dissipation
theorem. The equilibrium bond length was set to $\ell_0 =1.6$. The
simulations were done with $k_s = 1000$.

We performed in general $10^6$ integration time steps followed by
$10^4$ attempted pivot moves. A pivot move is an attempt to rotate a
portion of the chain by a small random angle around a randomly chosen
bead. The MC part is done with the usual metropolis algorithm accepting
pivot moves with a probability $\exp\left(\frac{- \Delta
    U}{k_{B}T}\right)$.  This mixed MD/MC procedure was repeated
$10^3$ times until the configurations were equilibrated. Equilibration
was checked by starting from crumpled chains and fully extended chains
and verifying that the same values for radius of gyration and
correlation functions was obtained. We simulated double-stranded
ribbon chains of $2 \times 400$ monomers. The simulations were performed on
a CRAY $T3D$ with 128 processors allowing us to simulate 128 chains in
parallel.

\begin{figure}

\centerline{\epsfxsize 10.0cm \epsffile{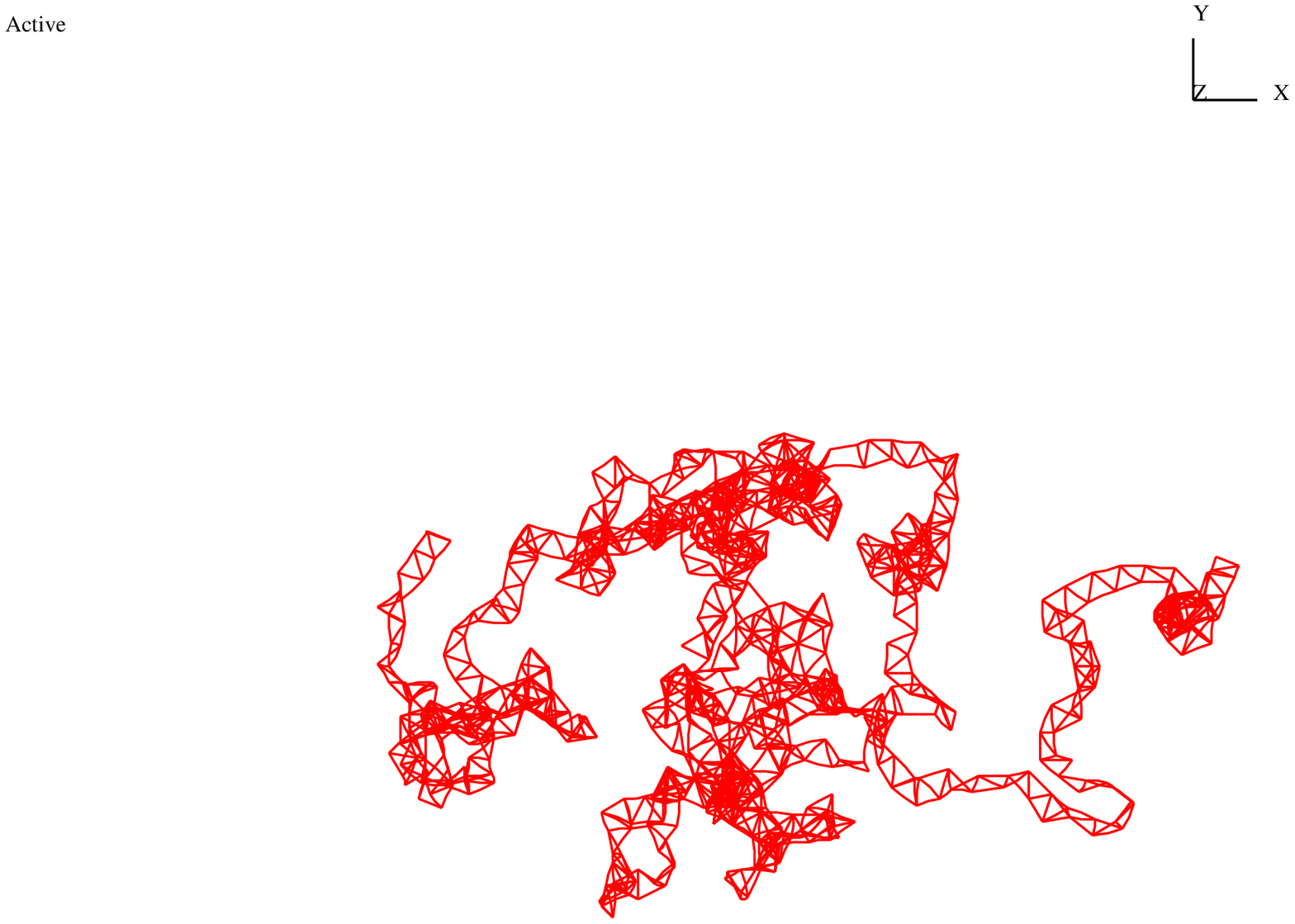}}
\caption{Typical conformations from MD/MC simulations of a ribbon made up
  of two chains of 400 monomers above $T_{\times}$, $k_b=1$.}
  \label{above}
\end{figure}

\begin{figure} 
\centerline{\epsfxsize 10.0cm \epsffile{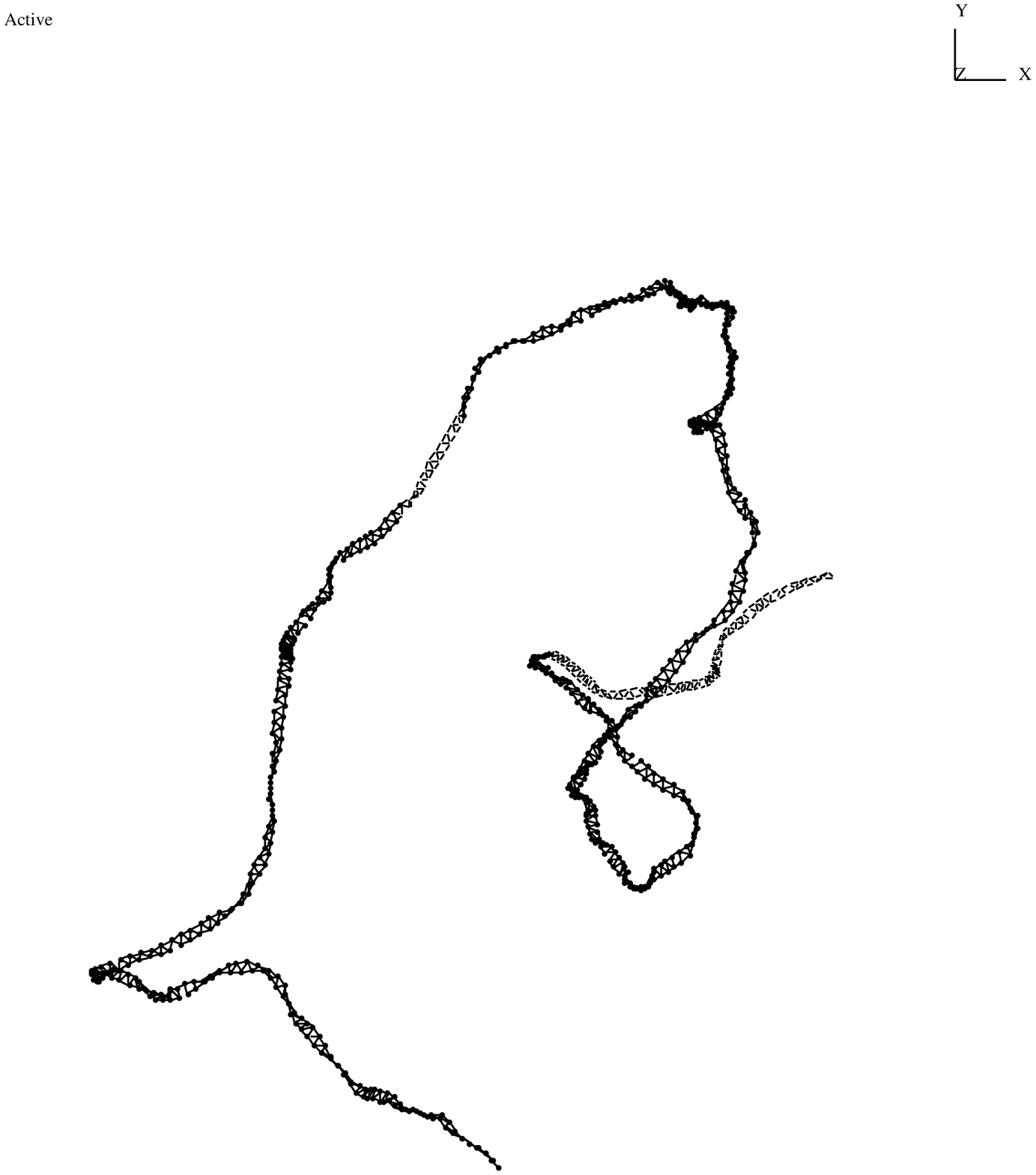}}
\caption{Typical conformations from MD/MC simulations of a ribbon made up
  of two chains of 400 monomers near $T_{\times}$, $k_b=10$.} \label{near}
\end{figure}

Typical equilibrated polymer configurations, shown here in 
Figs.~\ref{above},~\ref{near},~\ref{below}, suggest that
at low temperatures the polymer can be viewed as a collection of long,
twisted (straight) rods that are connected by short, highly curved
sections of chain which we call ``kinks'', as opposed to a smooth
worm-like conformation. This structure melts at higher temperatures.

We plot the $\langle {\bf b}(s) \cdot {\bf b}(0)\rangle$ correlation
function from the simulation in Fig.~\ref{bond_corr}. For $T> T_\times
$ we obtain simple exponential decay but for $T<T_\times$ we see an
oscillation in the correlation function in agreement with
Eq.(\ref{nn}).

We plot the $\langle {\bf t}(s) \cdot {\bf t}(0)\rangle$ correlation function from the
simulation in Fig.~\ref{tan_corr}.  We see the signature exponential
decay of the correlation function of worm-like chains from which we
can estimate the effective persistence length. The estimated
persistence lengths are $L_p =2.99 \pm 0.01 ,25.0 \pm 0.005 ,179.0 \pm
0.001$, correspondingly.

\section{Kink-Rod Structure}    \label{sKink}

\begin{figure}  
\centerline{\epsfxsize 10.0cm \epsffile{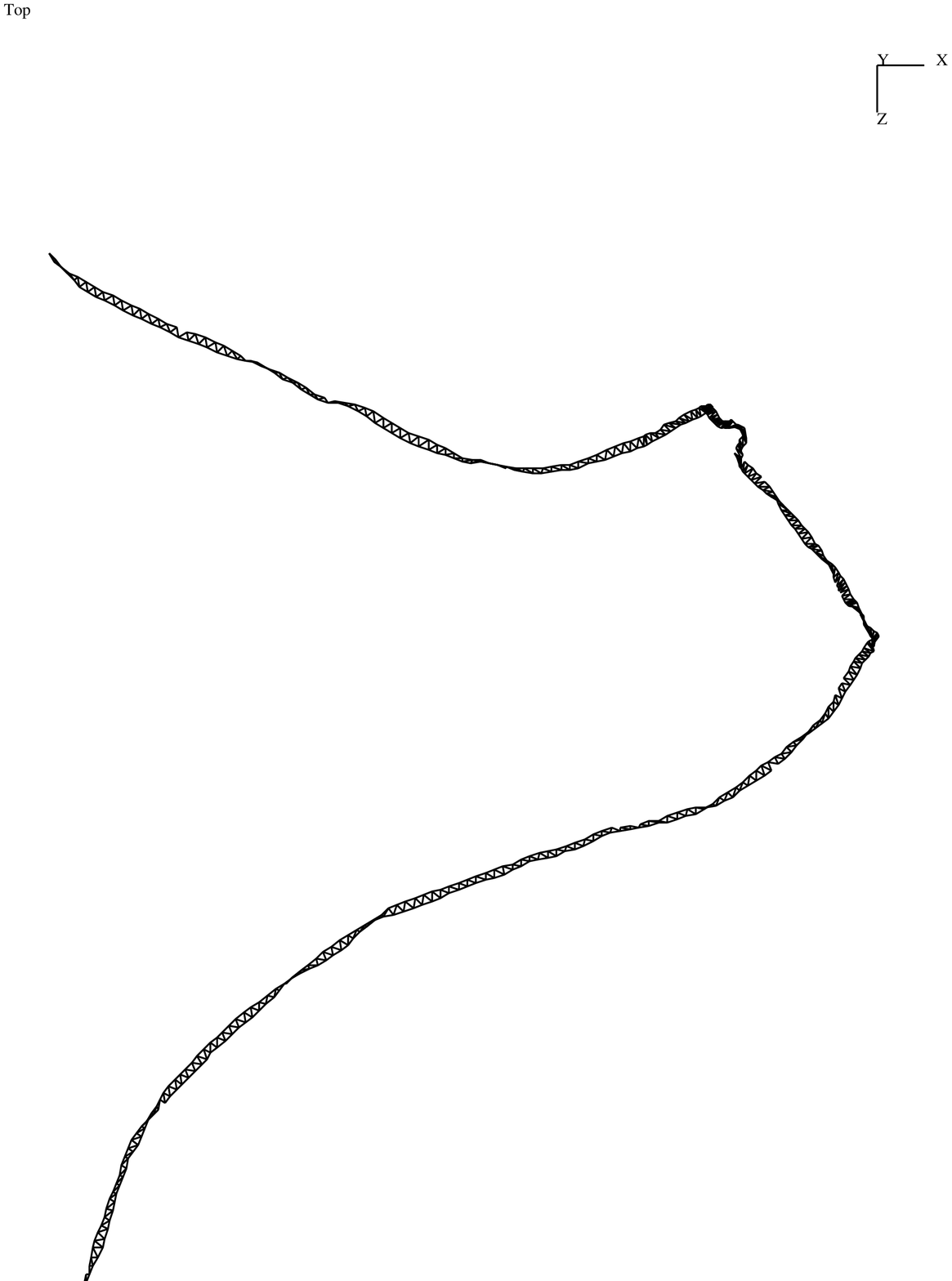}}
\caption{Typical conformations from MD/MC simulations of a ribbon made up
  of two chains of 400 monomers below $T_{\times}$, $k_b=100$.}
\label{below}\end{figure}

\begin{figure} 
\centerline{\epsfxsize 10.0cm \epsffile{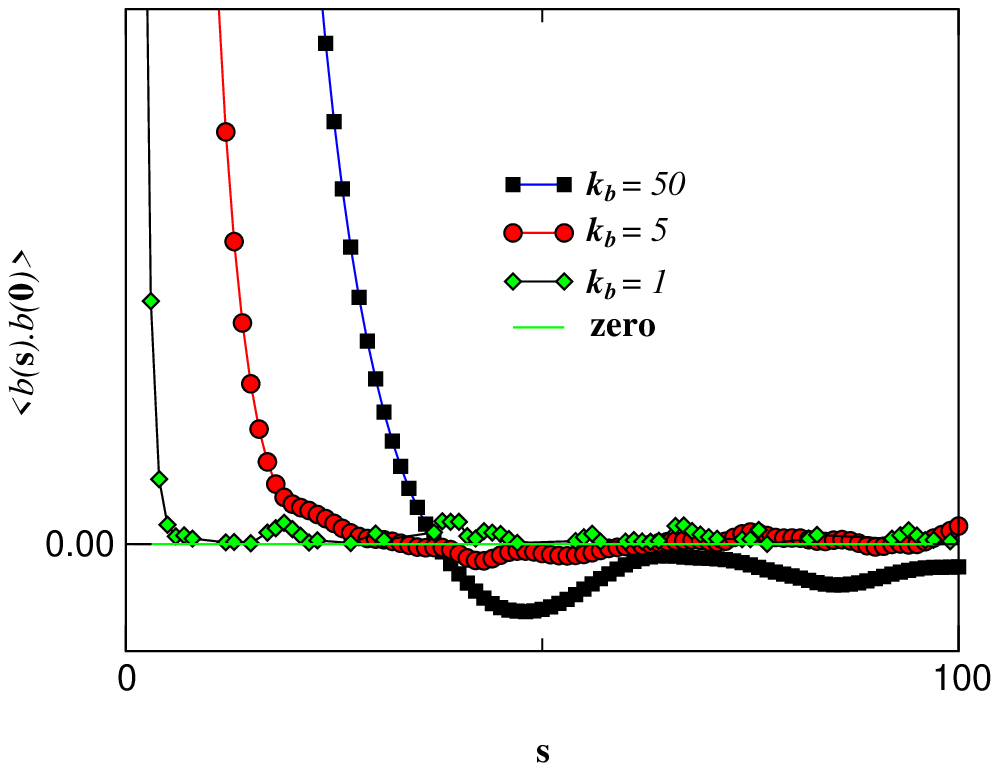}}
\caption{The $\langle {\bf b}(s) \cdot {\bf b}(0)\rangle$ correlation function
  measured in the simulations for temperatures $k_b=1,5, 50$
  corresponding to $b^{2} >c,b^{2}\approx c \,\, \mbox{and} \,\, b^{2} < c$.
  The averages were done over $\sim 10^{4}$ statistically independent
  samples. The error bars are the size of the symbols.}
 \label{bond_corr}\end{figure}

\begin{figure}  
\centerline{\epsfxsize 10.0cm \epsffile{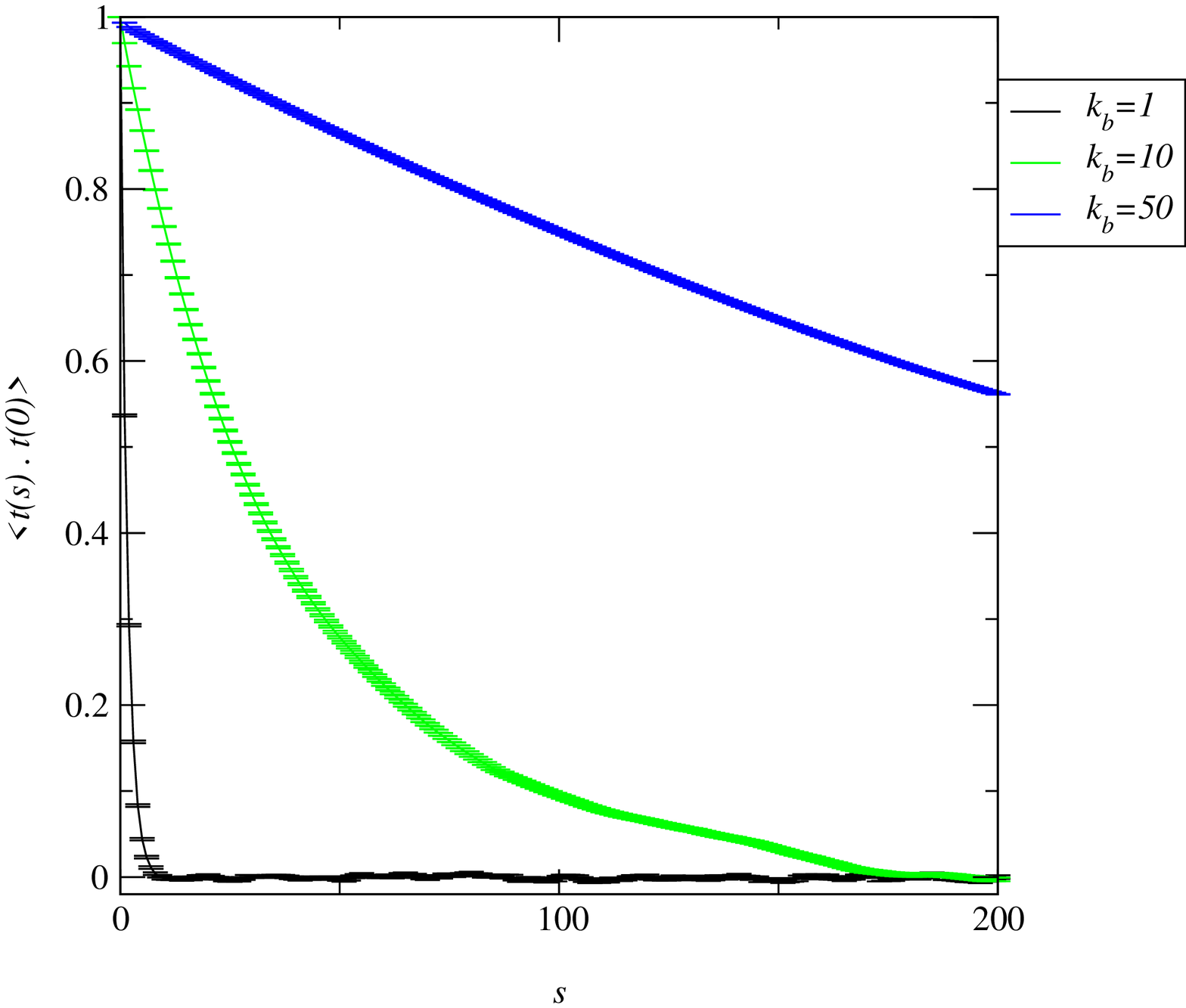}}
\caption{The $\langle {\bf t}(s) \cdot {\bf t}(0)\rangle$ correlation function
  measured in the simulations for temperatures $k_b=1,10,50$
  corresponding to $b^{2} >c,b^{2}\approx c \,\, \mbox{and} \,\, b^{2} < c$.
  The averages were done over $\sim 10^{4}$ statistically independent
  samples. The error bars are the size of the symbols.}
\label{tan_corr}\end{figure}

We show typical equilibrated conformations above, near, and below
$T_{\times}$ in Figs.~\ref{above},~\ref{near},~\ref{below}. The snapshots
of the polymer configurations suggest that at low temperatures the
polymer can be viewed as a collection of hard (straight) twisted rods
that are connected by some kinks. This picture can be accounted
for using a simple argument. We can model our system of two
semiflexible polymers subject to the constraint of constant
separation, as a semiflexible ribbon, i.e. a semiflexible linear
object with anisotropic rigidities whose Hamiltonian reads
\begin{equation}
{\cal H}_{\rm ani}= \frac{1}{2} \int \; \d s \; \sum_{{i,j}} \; \kappa_{ij} \;
\left(\frac{\d {\bf t}}{\d s}\right)_{i}\;
        \left(\frac{\d {\bf t}}{\d s}\right)_{j}, \label{kij}
\end{equation}
where $\kappa_{ij}=\kappa_{\parallel}\; b_i b_j+\kappa_{\perp}\; (\delta_{ij}-t_i t_j-b_i
b_j) $ determines the rigidity anisotropy of the ribbon, corresponding
to bending parallel or perpendicular to the bond-director field. The
ribbon structure would require $\kappa_{\parallel} \gg \kappa_{\perp}$. (To be
consistent with the hard constraint (see above) of constant separation
of the polymers, we should take the limit of infinite
$\kappa_{\parallel}$.) The partition function of a semiflexible ribbon in
the $\kappa_{\parallel} \to \infty$ limit can be written as

\begin{eqnarray}        \label{Zrib}
{\cal Z}_{\rm Rib}&=&\lim _{\kappa_{\parallel} \to \infty}
\int {\cal D} {\bf t}(s) {\cal D} {\bf b}(s) \exp\left\{-{\kappa_{\perp}
\over 2 k_B T} \int \d s \left(\frac{\d {\bf t}}{\d s}\right)^2
+ {\kappa_{\parallel}-\kappa_{\perp} \over 2 k_B T} 
\int \d s \left(\frac{\d {\bf t}}{\d s} \cdot {\bf b}\right)^2 
+{\cal H}'[{\bf t},{\bf b}]\right\} \\
&=&\int {\cal D} {\bf t}(s) {\cal D} {\bf b}(s) 
\delta\left\{\frac{\d {\bf t}}{\d s} \cdot {\bf b}\right\}
\exp\left\{-{\kappa_{\perp} \over 2 k_B T} 
\int \d s \left(\frac{\d {\bf t}}{\d s}\right)^2
+{\cal H}'[{\bf t},{\bf b}] \right\}, \nonumber 
\end{eqnarray} 
in which ${\cal H}'[{\bf t},{\bf b}]$ controls the dynamics of ${\bf b}$, 
and the functional delta-function enforces the constraint
\begin{equation}
\frac{\d {\bf t}(s)}{\d s} \cdot {\bf b}(s) = 0, \label{rodkink}
\end{equation}
to hold exactly at {\it every} point of the ribbon. Recalling $\d {\bf
  t}/{\d s}= H(s) \;{\bf n}$ from the Frenet-Seret
equations~\cite{geometry}, where $H(s)$ is the curvature at each point
and $ {\bf n}$ is the unit normal vector to the curve, we can write
the constraint as
\begin{equation}
H(s) \; {\bf n}(s) \cdot {\bf b}(s) =0. \label{Hnb}
\end{equation}
This constraint requires that at each point either $H(s)=0$, which
corresponds to a straight (rod-like) segment that can be twisted, or
${\bf n}(s) \cdot {\bf b}(s) =0$, which corresponds to a curved
(kink-like) region where the the bond-director is locked in to the
perpendicular direction to the curve normal, i.e., the {\it binormal}.

We expect the (core) length of the kink regions to be very short at
low temperatures, as observed in Fig.~\ref{below}.  We note that the
conformational entropy of the chain is due to the degrees of freedom
in the kink regions, whereas the twist entropy comes from the degrees
of freedom in the rod segments. The average separation between
neighbouring kinks is of the order of the persistence length. The
ribbon thus tends to keep the rod segments as long as possible to
maximally explore the twist degrees of freedom, while it can recover
the same conformational entropy as a wormlike chain from pivotal moves
in the kink regions. This explains the kink-rod structure in low
temperatures ($a \ll \ell_p$).  As the temperature increases, the
kinks get closer to each other, until at some temperature their
average separation becomes comparable to their size ($\ell_p \sim a$),
and the kink-rod pattern disappears.

This analysis can be understood in the context of the mean-field
$(\bf{b},\bf{t})$ model above. By observing that at low temperatures
${\ell}_{\rm BP} \ll {\ell}_{\rm TP}$, one can imagine
that there are roughly speaking rodlike (straight) segments of length
${\ell}_{\rm TP}$, each supporting a number of shorter segments
of length ${\ell}_{\rm BP}$ that are twisted, but de-correlated
with one other. One can see that ${\ell}_{\rm BP}$ is equal to
the length scale $\lambda$ at which the strands undergo conformational
fluctuations of the order of their separation: $a^2=<r^2>\equiv \int_{1/\lambda} \d
q/\ell_p q^4 \simeq \lambda^3/\ell_p$. Hence, segments of length
${\ell}_{\rm BP}$ are straight (for $a \ll \ell_p$). However, fluctuations of order
$a$ are sufficient to wash out the memory of twist. The
anti-correlation in fact comes from the frustration as explained above.

As the temperature is raised, the number of twisted rods in each
segment $N={\ell}_{\rm TP}/{\ell}_{\rm BP}$ decreases very
quickly, until it saturates to unity at $T=T_{\times}$. For higher
temperatures the mechanism changes, and the bond correlations are cut
off by the tangent fluctuations. Hence, the short-range twist order
does not survive anymore. All the main features of the above picture
have also been observed in the simulation.  

A kink-rod structure similar to the one discussed here has indeed been 
observed in experiments done on actin filaments \cite{Janmey}.
Actin is a charged polymer, and the mutual electrostatic repulsion of its
different segments plays a major role in its structural stiffness.
It is well known, however, that the introduction of {\it multivalent}
counterions (ions of opposite charge that are necessary to neutralise
the solution) can reduce the electrostatic repulsion, and even lead
to attraction between like charged polymers, or different like charged segments
of a same polymer \cite{GKL}.
In a recent experiment, Tang et al \cite{Janmey} used fluorescence microscopy
techniques to image {\it condensed} (or collapsed) actin bundles
that are formed due to the presence of multivalent counterions.
Snapshots of the bundles, showing their typical conformations, are shown 
in Fig.~\ref{actin}.
A remarkable feature in the observation was the presence of sharp
corners which connect relatively straight segments of the actin bundles,
as can be seen in Fig.~\ref{actin}. Tang et al \cite{Janmey} observed
that this feature is present only when the bundle is made up of two or
more filaments of actin, and is absent when there is only a single filament
in the condensate.

It is plausible to assume that the observed kink-rod structure can be accounted
for by similar arguments to the one developed above. The only difference is
the fact that the structure in the experimental case is a ring, as opposed 
to a chain with free ends. However, we do not expect this constraint to affect 
the argument, because the inherent competition between twist and bend
degrees of freedom is local. Of course, more experimental efforts are needed
to rule out other possible scenarios for the formation of the kinks, such as
defects in the packing of more than one filament, sequence disorder, or 
metastable effects due to the dynamics of the collapse.

\section{Mean Field Theory: Confined Chains}    \label{sConfine}

In this section, we study the effect of confinement on double-stranded
semiflexible polymers.  We confine the double-stranded polymer in a
$d_{\perp}$-dimensional subspace to a box of size $R$, while leaving
it free in the remaining $d_{\parallel}=d-d_{\perp}$ dimensions.

In terms of the tangent to the mid-curve ${\bf t}=\d {\bf r}/\d s$, and
the bond-director ${\bf b}$, the total Hamiltonian of the system can
now be written as
\begin{eqnarray} \frac{{\cal H}_{\rm conf}}{k_B T}& =&
\frac{{\cal H}+{\cal H}_m}{k_B T} + \frac{g}{\ell_{p}^3} \int \d s \;
 {\bf r}_{\perp}^2,\label{Htbconf}
\end{eqnarray}
where $g$ is a dimensionless constant in addition to $b$, $c$, $e$
defined in Section \ref{sFree}. The
constants will be determined self-consistently by demanding that the
relevant constraints hold on average: inextensibility of the
individual strands, constant separation between the chains, the
bond-director being normal to the strands, and confinement of the
polymer to a box of size $R$ in $d_{\perp}$ dimensions ($\left<{\bf
    r}_{\perp}^2(s) \right>=R^2$).  This final constraint is valid as
long as the chain length is much larger than $R$.  We obtain the
following set of equations for $b$, $c$, $e$, and $g$:
\begin{equation}        
        \left\{ \begin{array}{l}
        \sqrt{c}\frac{a^2}{4\ell_p^2}+\sqrt{g}\frac{R^2}{\ell_p^2}
        +\frac{d_{\parallel}}{4\sqrt{2b}}=1 \\ \\
        c(b+\sqrt{c})=\frac{d^2\ell_p^4}{2a^4}\nonumber \\ \\
        g(b+\sqrt{g})=\frac{d_{\perp}^2\ell_p^4}{32R^4}\nonumber \\\\
        e=0
        \end{array} \right. .\label{bcg}
\end{equation}
Although the above nonlinear set of equations are very
difficult to solve, we can get the behaviour of the solutions by
looking at the asymptotics in the limiting cases, as summarised in
Table \ref{asymp}. The full solutions are in fact smooth
interpolations between the asymptotics. We solved the equations
numerically for the experimentally relevant case; $d_{\perp}=1$ and
$d=3$.

Having determined the constants self-consistently, we can calculate
the correlation functions. For the tangent--tangent correlation in the
parallel direction one obtains
\begin{equation}
\left<{\bf t}_{\parallel}(s) \cdot {\bf t}_{\parallel}(0)\right>
=\frac{d_{\parallel}}{4 \sqrt{2 b}}
\exp\left(-\sqrt{2 b} \;\frac{s}{\ell_p}\right),\label{t11t11}
\end{equation}
whereas for the perpendicular direction one obtains
\begin{eqnarray}
\left<{\bf t}_{\perp}(s) \cdot {\bf t}_{\perp}(0)\right>
&=&\frac{d_{\perp}}{8 \sqrt{b^2-g}} 
\left( (b+\sqrt{b^2-g})^{1/2}
\exp\left[-(b+\sqrt{b^2-g})^{1/2}\;\frac{s}{\ell_p}\right]\right. \nonumber \\
&&-\left.(b-\sqrt{b^2-g})^{1/2}
\exp\left[-(b-\sqrt{b^2-g})^{1/2}\;\frac{s}{\ell_p}\right]
\right).\label{t+t+}
\end{eqnarray}
Similarly, for the bond-director field it yields
\begin{eqnarray}
\left<{\bf b}(s) \cdot {\bf b}(0)\right>=
\frac{d \ell_{p}^2}{2 a^2 \sqrt{b^2-c}}\left(
\frac{\exp\left[-(b-\sqrt{b^2-c})^{1/2}\;\frac{s}{\ell_p}\right]}{(b-\sqrt{b^2-c
})^{1/2}}
-\frac{\exp\left[-(b+\sqrt{b^2-c})^{1/2}\;\frac{s}{\ell_p}\right]}
{(b+\sqrt{b^2-c})^{1/2}} \right),\label{bbconf}
\end{eqnarray}
while the rest of the two-point functions (the cross terms) are zero.
The parallel component of the tangent-director correlation function
decays purely exponentially. However, the correlation function of the
perpendicular component of the tangent-director field, as well as that
of the bond-director field, develop a crossover from purely
exponential decay for $b^2 > g$ and $b^2 > c$, to oscillatory decay
for $b^2 < g$ and $b^2 < c$, respectively. The phase diagram of the
system in the space of dimensionless parameters $R/\ell_p$ and
$a/\ell_p$, is shown in Fig.\ref{fig2}.  The boundaries between
different regions are obtained from solutions of Eq.(\ref{bcg}).

It is instructive to examine the perpendicular component of the
tangent-director correlation function in the limiting case $b^2 \ll g$
which corresponds to $R/\ell_p \ll 1$ (see Table \ref{asymp}). 
Using the asymptotic forms for $b$ and $g$, one obtains
\begin{equation}
\left<{\bf t}_{\perp}(s) \cdot {\bf t}_{\perp}(0)\right>
=\left({d_{\perp}^2 \over 8\sqrt{2}} {R^2 \over \ell_p^2}\right)^{1/3}
\times \;\exp\left[-\left(\frac{d_{\perp}}{16 \ell_p R^2}\right)^{1/3} s\right]
 \times \; \sin\left[\left(\frac{d_{\perp}}{16 \ell_p R^2}\right)^{1/3} s+{\pi 
\over 4}\right].
 \label{t+t+asymp} 
\end{equation} 
The effects of confinement are best seen in this limiting expression.
The persistence length of the polymer in the confined directions is
reduced to $\ell_{\perp} \sim (\ell_p R^2)^{1/3}$. This `deflection
length'~\cite{Odijk} is in fact the length at which roughening of a
semiflexible chain of bare persistence length $\ell_p$ becomes
comparable to the confinement size (or separation of the confining
walls), $R$: $R^2=<r_{\perp}^2>=\int_{1/\ell_{\perp}} \d q/\ell_p
q^4$. In other words, the presence of the boundaries provides another
competing mechanism to cut off tangent correlations in the directions
of confinement. Moreover, the oscillatory form of the correlation
function with a period $\lambda = \ell_{\perp} \sim (\ell_p
R^2)^{1/3}$ implies a sinusoidal packing of the polymer in
the confining cavity, where again the size of the oriented segments
(the period of the oscillations) is set by the walls cutting off the
roughness of the polymer. As seen in Sec.\ref{sKink}, 
a semiflexible ribbon develops a kink-rod
structure at finite temperatures (due the strong 
anisotropy in rigidity), in which rod-like segments (about a 
persistence length
long) are connected by rather sharp kinks with a core size of the
order of the diameter of the ribbon. If such a structure is
restricted to a size less than the bare persistence length, new kinks
have to be created at the confining walls to squeeze the ribbon into
the available space. This therefore leads to the compact zigzag
conformation of kinks and rods and the subsequent oscillatory tangent
correlations.

It is straight forward to calculate the free energy of the
system. We find,
\begin{eqnarray}
{F \over k_B T}={F_0(b) \over k_B T}+{L \over 2 \ell_p}
\left[(b+\sqrt{b^2-c})^{1/2}+(b-\sqrt{b^2-c})^{1/2}
+(b+\sqrt{b^2-g})^{1/2}+(b-\sqrt{b^2-g})^{1/2}\right].\label{free1}
\end{eqnarray}
It is interesting to examine the limiting behaviour of the free energy
as a function of $R$. We obtain
\begin{equation}
{F(R) \over k_B T} \simeq {L \over 2 \ell_p} \left\{\begin{array}{cl}
\left({d_{\perp}/2}\right)^{1/3}
\left({\ell_p/R}\right)^{2/3}, & {\rm  for}\;\;{R/\ell_p} \ll 1  \\ \\ 
\left({\alpha \;d_{\perp}/d^2}\right)
\left({\ell_p/ R}\right)^{2}, & {\rm  for}\;\;{R /\ell_p} \gg 1  
\end{array} \right., \label{free2}
\end{equation}
where $\alpha$ is a smoothly varying numerical coefficient ranging
from $\alpha=4$ for $a /\ell_p \ll 1$, to $\alpha=1$ for $a /\ell_p
\gg 1$. The free energy, interestingly, interpolates between the
steric repulsion of non self-avoiding flexible polymers confined to $R$
\cite{deGennes}, which has $1/R^2$ behaviour, to Helfrich undulation
free energy of stiff polymers confined to $R$ \cite{Odijk,Helfrich}, which
has $1/R^{2/3}$ behaviour~\cite{undul}. 

Using the above results, we now analyse some recent experiments by 
Ott et al~\cite{Ott} who measure
the persistence length of actin filaments confined between microscope
slides.  The separation of the slides was of the order of $R = 1 \mu
m$ and they found a persistence length (assuming a two dimensional
worm-like chain) of $L_p^{0} = 16.7 \mu m$. These results are in the
regime addressed by our model. 
For very small {\it but}
finite $R$ the chain fluctuates between the two plates.  
>From the analogue of Eq.(\ref{t+t+asymp}) calculated for single-stranded 
semiflexible polymers, we find $\langle t_{\perp}^2 \rangle = 
({R/L_p^{(0)}})^{2/3}/2$.
Therefore one must include the fluctuations perpendicular to the
confining plates in the calculation of the true persistence length.
This implies that on average the polymer makes an angle $\theta$ given
by $ \sin \theta = \sqrt{\langle t_{\perp}^2 \rangle}$ with the plates
where $\theta < 1$. One therefore has a corrected persistence length
$L_p^{\mbox{true}} \approx L_p^{(0)} / \cos \theta = L_p^{(0)}
/\sqrt{1- \langle t_{\perp}^2 \rangle}$. We therefore estimate for
those experiments a correction of approximately $4 \%$, i.e.
$L_p^{\mbox{true}} = 17.4 \mu m$.

\section{Conclusion}    \label{sConcl}

In conclusion, we have calculated the properties of a well-defined
model of a double-stranded semiflexible polymer and shown novel
non-trivial differences between the high, low {\it and} zero
temperature behaviour. At high $T$ we find normal worm-like chain (WLC) 
behaviour and at low $T$ we observe a novel kink-rod structure with 
short-range twist order whilst at $T=0$ we have a flat ribbon. 

In the analytical approach, the only
approximation we have made is the relaxing of local constraints to
global ones. Using a {\em systematic} $1/d$-expansion (see Appendix)  
we have shown that to calculate the 2-point correlation functions 
this is a valid approximation as higher order corrections  only change 
the values of parameters but do not change the analytic form of the 
functions. Extensive MD/MC simulations confirm the analytical results. 

We have also examined the effect of confinement on the 
behaviour of semiflexible double-stranded polymers, and found
four interesting regimes of the conformation and internal twist 
structure of these polymers, as summarised in Fig.~\ref{fig2}:
(A) Weak confinement and relatively short bonds lead to free wormlike
chain conformations with short-ranged twist anti-correlations, (B)
weak confinement and relatively long bonds give rise to free wormlike
chain conformations and twist disorder, (C) strong confinement and
relatively short bonds yield sinusoidal packing of the chains
and short-ranged twist anti-correlations, and finally (D) strong confinement
and relatively long bonds lead to sinusoidal packing of the chains
and twist disorder.

There are a number of advantages evident in our approach. First, we
introduce a microscopic model which remains true to the chemical
structure of many biomolecules. Second, our {\it approximate} method
of solving this model also lends itself to the analysis of the
fluctuations in the system and to study intermediate-scale behaviour
as well as the ground state (long length-scale) properties.  Finally,
this method could be easily extended to describe multi-stranded
objects.  We expect that the effect of an intrinsic twist changes the
ground state but does not change any of the conclusions of our
description,  although we expect it to make the effective persistence
length much higher. We hope to address such
questions in a subsequent publication.

\acknowledgments It is a great pleasure to acknowledge many
stimulating discussions with A. Ajdari, R. Ejtehadi, R. Everaers, E.
Frey, G. Grest, M. Kardar, K. Kremer, A. Maggs and M. P{\"u}tz.  The
financial support of the Max-Planck-Gesellschaft and the Training and
Mobility of Researchers programme of the European Union is gratefully
acknowledged.  This research was supported in part by EU grant 
FMBICT972699 and the National Science Foundation under Grants No.
PHY94-07194, and DMR-98-05833.

\appendix

\section{$1/d$-expansion}       \label{a1/d}

To justify the mean field approximations used in Secs. \ref{sFree} and
\ref{sConfine} above, we perform a systematic $1/d$-expansion that
allows us to implement the constraints in a controlled way. Our
approach is similar to the one successfully used by David and Guitter
to study the crumpling transition of crystalline membranes
\cite{David}.  For simplicity, we consider a more primitive model of
an elastic ribbon with the Hamiltonian
\begin{equation}
{{\cal H}_{\rm ribbon} \over k_B T}={A \over 2}
\int \d s \left({\d {\bf t} \over \d s}\right)^2
+{C \over 2} \int \d s \left[{\d {\bf b} \over \d s}
-{\bf t}\left({\bf t}\cdot{\d {\bf b} \over \d s}\right)\right]^2, \label{Hrod}
\end{equation}
in which the tangent and the bond-director fields are subject to the
following constraints:${\bf t}^2=1, {\bf b}^2=1, {\bf t} \cdot {\bf
b}=0$, and ${\bf b}\cdot (\d {\bf t}/\d s)=0$ (Eq.(\ref{rodkink})). A
similar calculation with the more realistic (and more complicated)
Hamiltonian of Eq.(\ref{Htn}) will lead to essentially the same
conclusions. The partition function of the ribbon can be calculated as
\begin{equation}
{\cal Z}=\int {\cal D}{\bf t} {\cal D}{\bf b}{\cal D}\lambda_1
{\cal D}\lambda_2{\cal D}\lambda_3{\cal D}\lambda_4 \; {\rm e}^
{-S[{\bf t},{\bf b}, \lambda_1,\lambda_2,\lambda_3,\lambda_4]},\label{Zribbon}
\end{equation}
with
\begin{eqnarray}        \label{Stbllll}
S &=& {A \over 2}
\int \d s \left({\d {\bf t} \over \d s}\right)^2
+{C \over 2} \int \d s \left({\d {\bf b} \over \d s}\right)^2 \nonumber \\ &&
+i \int \d s \left[\lambda_1(s)\left({\bf t}^2-1\right)
+\lambda_2(s)\left({\bf b}^2-1\right)+\lambda_3(s)\left({\bf t} \cdot {\bf b} \right)
+\lambda_4(s)\left({\bf t}\cdot{\d {\bf b} \over \d s}\right)\right],
\end{eqnarray}
in which $\{\lambda_\alpha(s)\}$ are the ``stress'' (Lagrange multiplier) fields
enforcing the constraints. The integrations over ${\bf t}$ and ${\bf
  b}$ are now Gaussian, and can be performed to yield
\begin{eqnarray} \label{Seffllll}
S_{\rm eff}[\lambda_i]={d \over 2} \ln \det 
\left[\begin{array}{cc}
- A \partial_s^2+2 i \lambda_1& i \lambda_3 +i \lambda_4 \partial_s\\ 
i \lambda_3 -i \partial_s \lambda_4& - C \partial_s^2+2 i \lambda_2 
\end{array} \right]-i d \int \d s (\lambda_1+\lambda_2).
\end{eqnarray}
Note that we have rescaled $A,C$, and
$\{\lambda_\alpha(s)\}$ by $d$.  Extremising the effective action corresponds to
the saddle point or mean field solution ($d=\infty$). The saddle point
equations yield $i \bar{\lambda}_1=1/(8 A), i
\bar{\lambda}_2=/(8 C)$, and $\bar{\lambda}_3=\bar{\lambda}_4=0$.

To proceed to the higher orders in $1/d$, we need to calculate the $\lambda$-propagators 
defined as
\begin{equation}        \label{Gll}
\left<\lambda_\alpha (s)\lambda_\beta(s') \right>\equiv \left.{\delta^2 S_{\rm eff}
\over \delta \lambda_\alpha (s) \delta \lambda_\beta (s')} \right|^{-1}_{\rm saddle-point}.
\end{equation}
A straightforward calculation then leads to
\begin{eqnarray}
\left<\tilde\lambda_1(q) \tilde\lambda_1(-q)\right>&=&{A \over 8 d} q^2 +O(1),\\
\left<\tilde\lambda_2(q) \tilde\lambda_2(-q)\right>&=&{C \over 8 d} q^2 +O(1),\\
\left<\tilde\lambda_3(q) \tilde\lambda_3(-q)\right>&=&{A \over 2 d} q^2 +O(1),\\
\left<\tilde\lambda_4(q) \tilde\lambda_4(-q)\right>&=&
{A+C \over 2 d} q^2 +O(1/q^2),\\
\left<\tilde\lambda_3(q) \tilde\lambda_4(-q)\right>&=&{A \over 2 d} (-i q)+O(1/q),
\end{eqnarray}
while all the others are zero. Note that we have only kept the large momentum
limit, since we are interested in the local (short distance) behaviour of the
Lagrange multipliers \cite{David}. 

The above $\lambda$-propagators, the correlators for the tangent and the
bond fields
\begin{eqnarray}
\left<\tilde{t}_i(q) \tilde{t}_j(-q)\right>&=& 
{\delta_{ij}\over A q^2+1/(4 A)},\\
\left<\tilde{b}_i(q) \tilde{b}_j(-q)\right>&=& 
{\delta_{ij}\over C q^2+1/(4 C)},
\end{eqnarray}
and the three-point vertices
\begin{eqnarray}
\left<\tilde{t}_i(q) \tilde{t}_j(q') \tilde\lambda_1(-q-q')\right>&=& 2 i \delta_{ij},\\
\left<\tilde{b}_i(q) \tilde{b}_j(q') \tilde\lambda_2(-q-q')\right>&=& 2 i \delta_{ij},\\
\left<\tilde{t}_i(q) \tilde{b}_j(q') \tilde\lambda_3(-q-q')\right>&=& i \delta_{ij},\\
\left<\tilde{t}_i(q) \tilde{b}_j(q') \tilde\lambda_4(-q-q')\right>&=& -q' \delta_{ij},\\
\end{eqnarray}
as read from Eq.(\ref{Stbllll}), could now be used to construct diagrammatic expansions.
Examining the 2-point correlation functions for ${\bf t}$ and ${\bf b}$, we then find
that the perturbative expansions are well-behaved (not singular), and only correct
the numerical values of the coupling constants by finite amounts at each order.
We thus conclude that the mean field behaviour corresponding to the saddle point
approximation is qualitatively valid.

\begin{figure}
%
%
\caption{Snapshots of condensed actin rings due to the presence of multivalent
  counterions, from the experiment by Tang et al
  \protect\cite{Janmey}. The kink-rod structure is manifest, in
  bundles with two or more filaments \protect\cite{Janmey}.}
\label{actin}
\end{figure}
\begin{figure}
\centerline{\epsfxsize 10cm {\epsffile{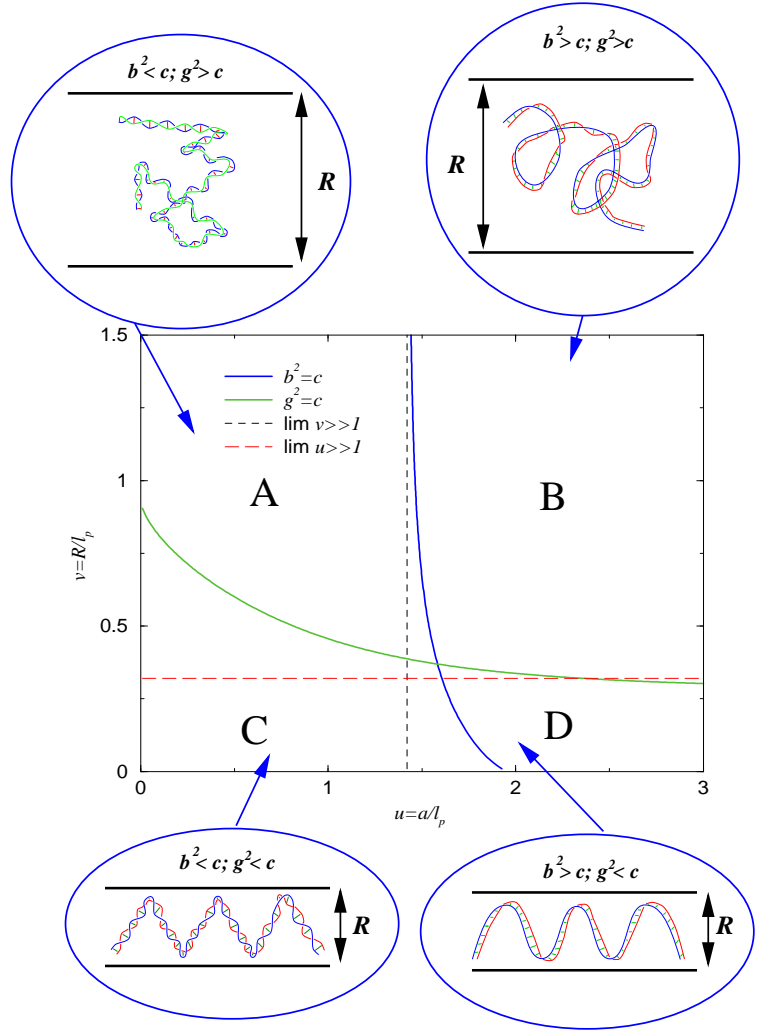}}}
\caption{The phase diagram as a function of $v=R/\ell_p$ and $u=a/\ell_p$ 
calculated numerically using Eq.(\ref{bcg}). We work in three dimensions $d=3$ 
and with confinement in one direction, $d_{\perp}=1$.}
\label{fig2}
\end{figure}
%
%
%
\begin{table}
\caption{The limiting behaviour of the confined polymer: Asymptotic behaviour of the 
solutions of Eq.(\ref{bcg}).}
\begin{center}
\begin{tabular}{cccccc} \hline 
 & & & $b$ & $c$ & $g$  \\ \hline \hline
$(1)$ & ${R \over \ell_p}\ll 1$ & ${a \over \ell_p}\ll 1$ 
&$\frac{d_{\parallel}^2}{32}$ 
& $\left(\frac{d}{\sqrt{2}}\frac{\ell_p^2}{a^2}\right)^{4/3}$ 
& $\left(\frac{d_{\perp}}{4\sqrt{2}}\frac{\ell_p^2}{R^2}\right)^{4/3}$ \\ \hline
(2) & ${R \over \ell_p}\gg 1$ & ${a \over \ell_p}\ll 1$ &$\frac{d^2}{32}$
& $\left(\frac{d}{\sqrt{2}}\frac{\ell_p^2}{a^2}\right)^{4/3}$ 
& ${d_{\perp}^2 \over d^2} {\ell_p^4 \over R^4}$ \\ \hline
(3) & ${R \over \ell_p}\ll 1$ & ${a \over \ell_p}\gg 1$ 
&$\frac{(d+d_{\parallel})^2}{32}$
& ${16 d^2 \over (d+d_{\parallel})^2} {\ell_p^4 \over a^4}$ 
& $\left(\frac{d_{\perp}}{4\sqrt{2}}\frac{\ell_p^2}{R^2}\right)^{4/3}$ \\ \hline
(4) & ${R \over \ell_p}\gg 1$ & ${a \over \ell_p}\gg 1$ &$\frac{d^2}{8}$
& ${4 \ell_p^4 \over a^4}$
& ${d_{\perp}^2 \over 4 d^2} {\ell_p^4 \over R^4}$ \\ \hline
\end{tabular}
\end{center}
\label{asymp}
\end{table}


\begin{references}
\bibitem{like}
Note that we have changed our sign convention from ~\cite{LGK} in
defining twists.
\bibitem{undul} In general, the entropic undulation free energy for a
  $D$-dimensional rigid manifold consists of contributions of a $k_B
  T$ per contact, i.e. $F/L^D \sim k_B T/\xi^D$, where $\xi$, the
  Odijk deflection length, is the length scale at which the roughness
  of the rigid manifold causes the first contact: $R^2=<r_{\perp}^2>
  \sim \int_{1/\xi} \d^D q/q^4 \sim \xi^{(4-D)}$.  Replacing $\xi \sim
  R^{2/(4-D)}$ in the former expression yields a Helfrich undulation
  free energy that goes like $1/R^{2D/(4-D)}$.  For rigid membranes
  (D=2) it gives $1/R^2$, whereas for stiff polymers (D=1) it gives
  $1/R^{2/3}$.
\end{references}
\end{document}